\documentclass[12pt]{iopart}

\usepackage{amssymb}
\usepackage{color}
\usepackage{graphicx}
\usepackage{dcolumn}
\usepackage{bm}
\usepackage{array}
\usepackage{subfig}
\newcommand{\PreserveBackslash}[1]{\let\temp=\\#1\let\\=\temp}
\newcolumntype{C}[1]{>{\PreserveBackslash\centering}p{#1}}
\newcolumntype{R}[1]{>{\PreserveBackslash\raggedleft}p{#1}}
\newcolumntype{L}[1]{>{\PreserveBackslash\raggedright}p{#1}}

\begin{document}

\title{Uncovering missing links with cold ends}

\author{Yu-Xiao Zhu$^1$, Linyuan L\"u$^2$, Qian-Ming Zhang$^{1,3}$, Tao Zhou$^{1,3*}$}

\address{$^1$Web Sciences Center, University of Electronic Science and Technology of China, Chengdu 611731, P. R. China\\
$^2$Department of Physics, University of Fribourg, Chemin du Muse,
CH-1700 Fribourg, Switzerland \\
$^3$Beijing Computational Science Research Center, Beijing 100089, P. R. China}

\ead{zhutou@ustc.edu}

\begin{abstract}
To evaluate the performance of prediction of missing links, the
known data are randomly divided into two parts, the training set and
the probe set. We argue that this straightforward and standard
method may lead to terrible bias, since in real biological and
information networks, missing links are more likely to be links
connecting low-degree nodes. We therefore study how to uncover
missing links with low-degree nodes, namely links in the probe set
are of lower degree products than a random sampling. Experimental
analysis on ten local similarity indices and four disparate real
networks reveals a surprising result that the Leicht-Holme-Newman
index [E. A. Leicht, P. Holme, and M. E. J. Newman, Phys. Rev. E
{\bf 73}, 026120 (2006)] performs the best, although it was known to
be one of the worst indices if the probe set is a random sampling of
all links. We further propose an parameter-dependent index, which
considerably improves the prediction accuracy. Finally, we show the relevance of the proposed index on three real sampling methods.
\end{abstract}


\maketitle
\section{Introduction}

Many social, biological, and information systems can be well
described by networks, where nodes represent individuals and links
denote the relations or interactions between nodes. The study of
complex networks has therefore become a common focus of many
branches of science. A fundamental tool for network analysis is the
so-called \emph{link prediction}, which attempts to estimate the
likelihood of the existence of a link between two nodes, based on
observed links and the attributes of nodes
\cite{LPsurvey,Getoor2005}.

In many biological networks, such as food webs, protein-protein
interaction networks and metabolic networks, whether a link between
two nodes exists must be demonstrated by field and/or laboratory
experiments, which are usually very costly. Our knowledge of these
networks is very limited, for example, 80\% of the molecular
interactions in cells of Yeast \cite{Yu2008} and 99.7\% of human
\cite{Stumpf2008,Amaral2008} are still unknown. Instead of blindly
checking all possible interactions, to predict based on known
interactions and focus on those links most likely to exist can
sharply reduce the experimental costs if the predictions are
accurate enough. Social network analysis also comes up against the
missing data problem \cite{Kossinets2006,Neal2008}, where link
prediction algorithms may play a role. In addition, the data in
constructing biological and social networks may contain inaccurate
information, resulting in spurious links
\cite{Mering2002,Butts2003}. Link prediction algorithms can be
applied to identify these spurious links \cite{Guimera2009}. Besides
helping in analyzing networks with missing data, the link prediction
algorithms can be used to predict the links that may appear in the
future of evolving networks. For example, in online social networks,
very likely but not yet existent links can be recommended as
promising friendships, which can help users in finding new friends
and thus enhance their loyalties to the web sites. Other
applications of link prediction include the evaluation of network
evolving models \cite{LiuHK}, the classification of partially
labeled networks \cite{Zhang2010}, and so on (see the review article
\cite{LPsurvey} for the detailed discussion on real applications).

To evaluate the algorithmic performance, the data set is divided
into two parts: the training set is treated as known information
while the probe set is used to estimate the prediction accuracy. To
our knowledge, the datasets are always divided completely randomly. This is of course the most straightforward way, and it seems a very
fair method without any statistical bias. However, this
straightforward and standard method may lead to terrible bias, since
in real biological and information networks, missing links are more
likely to be links connecting low-degree nodes. For example, the
known structure of the World Wide Web is just a sampling, where the
hyperlinks from popular web pages have higher probability to be
uncovered. In contrast, hyperlinks from unbeknown web pages are
probably lost. Actually, in common sense, interaction between two
significant proteins, hyperlink between two well-known web pages and
relationship between two famous persons are of less probability to
be missed. Accordingly, in this article, we study how to uncover
missing links with low-degree nodes. That is to say, we divide the
data set into two parts and make the links in the probe set less
popular (i.e., of less degree products) than the links in the
training set. Experimental analysis on ten local similarity indices
and four disparate real networks reveals a surprising result that
the Leicht-Holme-Newman (LHN) index \cite{LHN} performs the best,
although it was known to be one of the worst indices if the probe
set is a random sampling of all links \cite{ZhouEPJB2009}. We
further propose an parameter-dependent index, which considerably
improves the prediction accuracy. Finally, we show the relevance of the proposed index on three real sampling methods.

This article is organized as follows. In the next section, we will
clearly define the problem of link prediction, describe the
standard process to evaluate the prediction accuracy, introduce the
state-of-the-art local indices for node similarity and how to sample
less popular links for the probe set. Experimental results for the
traditional sampling method and the proposed method are presented in
Section III. In Section IV, we will propose an improved index which
performs even better than the LHN index. In Section V, we will introduce three mainstream sampling methods, and test the improved index on the corresponding sampled networks. Finally, we summarize our results in Section VI.

\section{Problem Description}
\subsection{Link Prediction: Problem and Evaluation}
Given an undirected network $G(V,E)$, where $V$ and $E$ are the sets
of nodes and links respectively. The multiple links and
self-connections are not allowed. Denote by $U$ the universal set
containing all $\frac{|V|(|V|-1)}{2}$ possible links, where $|V|$
denotes the number of elements in set $V$. Then, the set of
nonexistent links is $U\backslash E$, in which there are some missing links, namely the existed
yet unknown links or those that will form in the future. The task of
link prediction is to uncover these links. Each node pair $x$ and
$y$ will be assigned a score $s_{xy}$ according to a given
prediction algorithm. The higher score, the higher probability that
this link exists. The score matrix $S$ is symmetry for $G$ is
undirected. All the nonexistent links are sorted in descending order
according to their scores, and the top-ranked links are most likely to
exist.

To test the algorithmic accuracy, the observed links $E$ are divided
into two groups: the training set $E^{T}$ is treated as known
information, while the probe set $E^{P}$ is used for testing and no
information therein is allowed to be used for prediction. Clearly,
$E = E^{T}\cup E^{P}$ and $E^{T}\cap E^{P} = \phi$. The accuracy of
prediction is quantified by a standard metric called AUC (short for
\emph{area under the receiver operating characteristic curve})
\cite{AUC}. Specifically, this metric can be interpreted as the
probability that a randomly chosen missing link (links in $E^{P}$)
has higher score than a randomly chosen nonexistent link (links in
$U\backslash E$). In the implementation, among $n$ times of
independent comparisons, if there are $n'$ times that the missing
link has higher score and $n''$ times the missing link and
nonexistent link have the same score, AUC is calculated by
$\frac{n'+0.5n''}{n}$. If all the scores are generated from an
independent and identical distribution, the accuracy should be about
0.5. Therefore, the degree to which the accuracy exceeds 0.5
indicates how much better the algorithm performs than pure chance.

\subsection{Similarity Index}

The simplest framework of link prediction is the
similarity-based algorithm, where each pair of nodes, $x$ and $y$,
is assigned a score $s_{xy}$, which is directly defined as the
similarity between them \cite{LPsurvey,Liben-Nowell,Lu2009}. All non-observed links are ranked according
to their scores, and the links connecting more similar nodes are
supposed to be of higher existence likelihoods. Owning to its
simplicity, the study on similarity-based algorithms is the
mainstream issue.

In this article, we adopt the simplest local similarity indices.
Zhou \emph{et al.} \cite{ZhouEPJB2009} have investigated the
performances of these ten local indices, including Common Neighbors
(CN), Salton index \cite{Salton1983}, Jaccard index \cite{Jaccard},
S{\o}rensen index \cite{Sorensen}, Hub Promoted index (HPI)
\cite{HPI}, Hub Depressed index (HDI) \cite{ZhouEPJB2009},
Leicht-Holme-Newman index (LHN) \cite{LHN}, Preferential Attachment
(PA) \cite{barabasi1999,PA}, Adamic-Adar Index (AA) \cite{AA} and
Resource Allocation index (RA) \cite{ZhouEPJB2009,Ou2007}. It was
shown that the RA index performs best, and LHN and PA indices
perform the worst. However, these results are obtained based on
random probe set division. In this article, we will compare the
performances of these ten indices on predicting the missing links
with low-degree nodes. Their mathematical expressions are shown in
Table~\ref{table1}. The detailed information on these ten indices
can be found in Refs. \cite{ZhouEPJB2009,LPsurvey}. Note that, the
above indices except PA, are all common-neighbor based. Therein
Salton index, Jaccard index, S{\o}rensen index, HPI, HDI and LHN are
different in the dominators which take into account the degrees of
the two endpoints of the predicted links, while AA and RA indices
consider the effects of their common neighbors' degrees.

\begin{table}
\renewcommand{\arraystretch}{1.5}
\caption{Mathematical expressions of ten local similarity indices.
Denote by $k_x$ and $\Gamma_x$ the degree of node $x$ and the set of
its neighbors. See Ref. \cite{ZhouEPJB2009} for details.}
\begin{center}
\begin{tabular}{ll}
  \hline\hline
    $s^{\texttt{CN}}_{xy}=|\Gamma_x\cap\Gamma_y|$ \quad \quad & $s^{\texttt{PA}}_{xy} = k_x\times k_y$\\

    $s^{\texttt{Salton}}_{xy}= \frac{|\Gamma_x\cap\Gamma_y|}{\sqrt{k_x\times k_y}}$ \quad \quad & $s^{\texttt{Jaccard}}_{xy}=\frac{|\Gamma_x\cap\Gamma_y|}{|\Gamma_x\cup\Gamma_y|}$ \\

   $s^{\texttt{S{\o}rensen}}_{xy} =
  \frac{|\Gamma_x\cap\Gamma_y|}{{k_x+k_y}}$ \quad \quad & $s^{\texttt{HPI}}_{xy}=\frac{|\Gamma_x\cap\Gamma_y|}{\min\{k_x,k_y\}}$\\

    $s^{\texttt{HDI}}_{xy}=\frac{|\Gamma_x\cap\Gamma_y|}{\max\{k_x,k_y\}}$ \quad \quad & $s^{\texttt{LHN}}_{xy} = \frac{|\Gamma_x\cap\Gamma_y|}{k_x\times k_y}$\\

    $s^{\texttt{AA}}_{xy}=\sum\limits_{z\in\Gamma_x\cap\Gamma_y}\frac{1}{\log{k_z}}$  \quad \quad & $s^{\texttt{RA}}_{xy}=\sum\limits_{z\in \Gamma_x\cap \Gamma_y}\frac{1}{k_z}$\\

  \hline\hline
    \end{tabular}\label{table1}
\end{center}
\end{table}

\subsection{Sampling for Probe Set}
Traditionally, the probe links are randomly selected from $E$,
namely each link has equal probability to be selected into probe set
(called \emph{random sampling}). In this way, the algorithmic
accuracy measured by AUC is actually an average prediction accuracy
of the probe links. However, the links may have different
predictabilities for their different roles in the network. Some
prediction algorithms may be good at predicting the links connecting
the high-degree nodes, while some are adept in the links connecting
the low-degree nodes. Therefore, in order to evaluate the
performance of different algorithms on different links, the dataset
should be divided with preference.

Motivated by evaluating the algorithm's performance on uncovering
the links with low-degree nodes, in this paper, we propose a preferential
partition method according to the link \emph{popularity} which is defined as:
\begin{equation}
\mathrm{pop}_{(x,y)} = (k_x-1) \times (k_y-1), \label{Eq1}
\end{equation}
where $k_x$ denotes the degree of node $x$. Clearly, links with
high-degree endpoints have higher popularities than those with
low-degree ends. Thus for a given network, the links whose
popularities are higher than the average popularity $\langle
\mathrm{pop}\rangle$ are popular links, and those with lower
popularities than $\langle \mathrm{pop}\rangle$ are unpopular links.
The detailed partition steps are as follows: (i) Calculate the
popularity score of each observed link according to Eq.~\ref{Eq1},
and rank these links in descending order based on their popularity
scores. (ii) Uniformly divide this list from down to up into $D$
groups respectively denoted by $E_{1},E_{2},...,E_{D}$. Clearly,
$E_{1}\cup E_{2}\cup ... \cup E_{D}=E$ and $E_{i}\cap
E_{j}=\phi,(i,j = 1,2,...D, i\neq j)$. The popularity of each link
in $E_{i}$ is no higher than that in $E_{j}$ if $i<j$. (iii) For
each subset $E_{i}$, we randomly choose half of the links therein to
constitute the probe set labeled by $E_{i}^{P}$. Then the rest links
(i.e., $E\backslash E_i^P$) constitute the corresponding training
set labeled by $E_i^T$. Denote by $\langle \mathrm{pop}\rangle_i$
the average popularity of the links in probe set $E_{i}^P$, we have
$\langle \mathrm{pop}\rangle_1<\langle
\mathrm{pop}\rangle_2<,\cdots,\langle \mathrm{pop}\rangle_D$.
$E_{1}^{P}$ consisting of the most unpopular links are called cold
probe set in this article. We design this sampling method for the convenience of theoretical analysis. However, this method is far different from real sampling methods. We will therefore test the relevance and validity of our main results on real sampling methods in Section 5.

\section{Experimental analysis}
\subsection{Data Description}
We consider four representative networks drawn from disparate
fields: (i) USAir: The network of US air transportation system,
which contains 332 airports and 2126 airlines \cite{USAir}. (ii)
NetScience: A network of coauthorships between scientists who are
themselves publishing on the topic of networks \cite{NS}. The
network contains 1589 scientists, 128 of which are isolated. Here we
consider the largest component that contains only 379 scientists.
(iii) C.elegans: The neural network of the nematode worm C.elegans,
in which an edge joins two neurons if they are connected by either a
synapse or a gap junction \cite{CC}. This network contains 297
neurons and 2148 links. (iv) Political Blogs: The network of US
political blogs \cite{PB}, the original links are directed, here we
treat them as undirected links. Table \ref{table2} summarizes the
basic topological features of these networks. Brief definitions of
the monitored topological measures can be found in the table
caption. For more details, one can see the review articles
\cite{data_rv1,data_rv2,data_rv3,data_rv4,data_rv5}.

\begin{table*}[htbp]
\begin{center}
\caption{The basic topological features of the giant components of
the four example networks. NS, CE and PB are the abbreviations for
NetScience, C.elegans and Political Blogs networks respectively.
$N=|V|$ and $M=|E|$ are the total number of nodes and links,
respectively. $C$ and $r$ are clustering coefficient \cite{CC} and
assortative coefficient \cite{assortative_cofficient}. $\langle
k\rangle$ is the average degree of network. $\langle d\rangle$ is
the average shortest distance between node pairs. $H$ denotes the
degree heterogeneity defined as $H=\frac{\langle
k^{2}\rangle}{\langle k\rangle^{2}}$.}
{\begin{tabular}{C{1.5cm}|C{1.0cm}C{1.1cm}C{1.1cm}C{1.1cm}C{1.0cm}C{0.9cm}C{0.9cm}}
\hline\hline
 DataSets &   $N$    &    $M$    &    $C$    &    $r$    &    $\langle k\rangle$    &    $\langle d\rangle$    &   $H$   \\
\hline
USAir &    332    &    2126   &   0.749   &   -0.208   &   12.81 &    2.46   &   3.46  \\
NS    &    379    &    914    &   0.798   &   -0.082   &    4.82 &    4.93   &   1.66  \\
CE    &    297    &   2148    &   0.308   &   -0.163   &   14.46 &   2.46   &   1.80  \\
PB    &    1222   &   16717   &   0.361   &   -0.221   &   27.36 &    2.51   &   2.97  \\
\hline\hline
\end{tabular}\label{table2}}
\end{center}
\end{table*}

\subsection{Results for Random Probe Set}
\begin{figure}[htbp]
\centering
\includegraphics[width=6.2in]{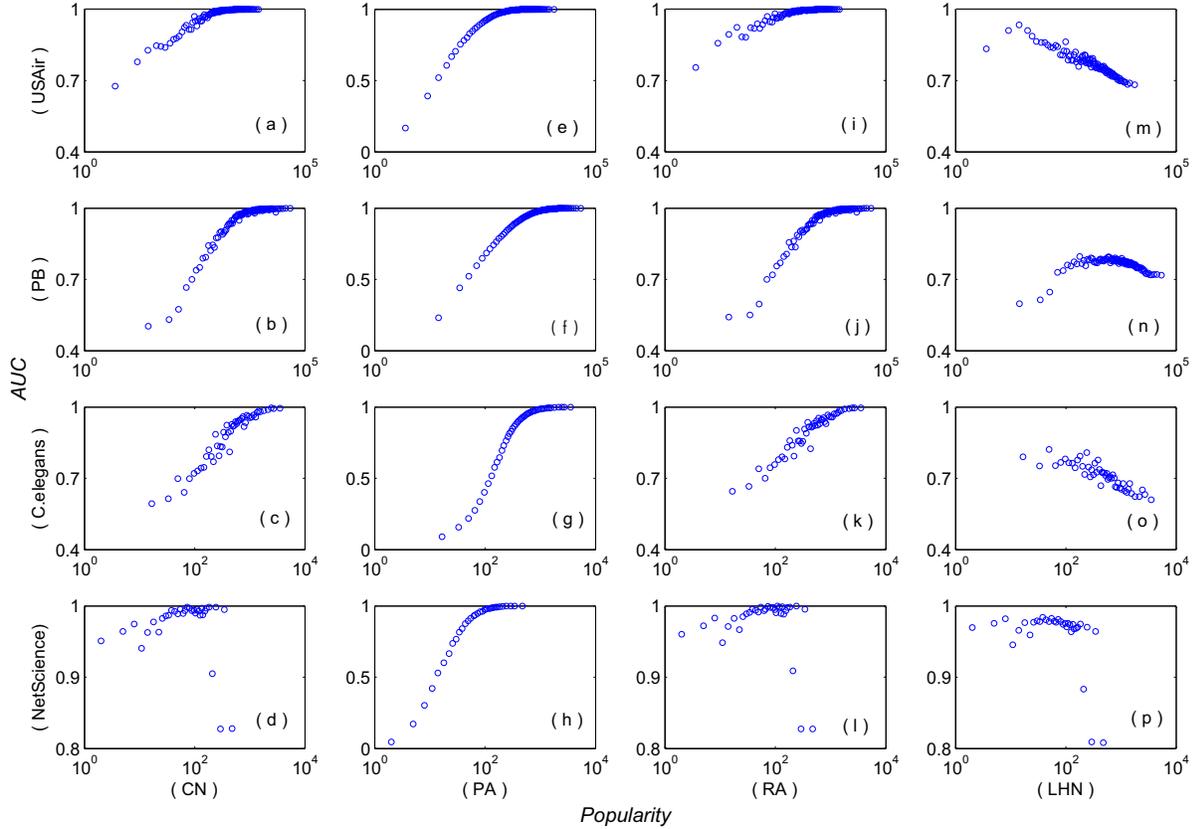}
\caption{The dependence of algorithmic accuracy on the popularity of
links. The results for four typical indices, namely CN, PA, RA and
LHN are shown respectively in four columns. Each subgraph is
obtained by averaging over 100 implementations with independently
random partitions into training set and probe set. The probe set
contains 5\% of observed links. Note that, the AUC value corresponds
to the average AUC of links with the same popularity. The statistics
are conducted with log-bin of popularity. \label{fig1}}
\end{figure}

As we have mentioned above, the mainstream method to prepare the
probe set is random sampling, namely all the links in $E^{P}$ are
randomly chosen from the whole link set $E$. In this way, AUC gives
the average performance on predicting the links in probe set. For
example, Zhou \emph{et al.} compared ten local similarity indices on
five real networks \cite{ZhouEPJB2009} with this randomly selected
probe set, and gave an overall evaluation measured by AUC. Instead
of obtaining a collective performance of the whole probe set, here
we investigate the algorithm's performance on each link. The
accuracy of one link is defined as the probability that this link
has higher score than that of one randomly chosen nonexistent link.
The dependence of four typical algorithms' accuracies on the
popularity of links is shown in Fig.~\ref{fig1}. Note that the
popularity of each link in probe set is calculated according to the
initial dataset, not the training set. Fig.~\ref{fig1}(a)-(l) show
that the AUC increases with the increasing of link popularity. This
indicates that the CN, PA and RA algorithms tend to give higher
accurate predictions on popular links, especially in USAir, PB and
C.elegans networks. In comparison, the LHN index, which has been
demonstrated to be a low accurate method in previous works, can give
higher accurate prediction on the unpopular links. The reason is LHN
is likely to assign higher score to the unpopular links by using
$k_x\times k_y$ as its dominator to depress the scores of popular
links.

We further investigate the average popularity of the top-$L$
predicted links of different algorithms. In principle, a link
prediction algorithm provides a descending ordered list of all
non-observed links (i.e., $U\backslash E^P$ in our experiment)
according to their scores, of which we only focus on the top-$L$
links. Then, the top-$L$ popularity is defined as the average
popularity of links among the top-$L$ places. A low top-$L$ average
popularity indicates that the algorithm tends to rank the missing
links connecting low-degree nodes at the top places. Table
\ref{top-L-popularity} shows the top-$L$ popularity of ten local
similarity indices on C.elegans network. For CN, PA, AA and RA
indices the top-100 popularity scores are extremely high, lager than
100, and the scores will decrease with the increasing of $L$. This
indicates that these four indices tend to rank the popular links at
the top places. For the other six indices, namely Salton, Jaccard,
S{\o}rensen, HPI, HDI and LHN, the top-100 popularity scores are
very low. Especially the score of LHN index is very small,
approximate to zero, and will increase with the increasing of $L$,
suggesting that the LHN index is likely to assign higher scores to
the links among whose two endpoints there is at least one node with
degree equal to 1. When $L$ is large, the overlap of two ranking
lists generated by two algorithms is very high, and thus leads to
similar top-$L$ popularity scores. This result further demonstrates
that LHN index is more competent to uncover the unpopular links,
especially the links with very low-degree nodes.

\begin{table*}[htbp]
\centering
\caption{Algorithmic novelty of all local similarity
indices on C.elegans, measured by the top-$L$ popularity. Sal, Jac and S{\o}r are the abbreviations for
Salton, Jaccard and S{\o}rensen methods respectively. Each
number is obtained by averaging 100 implementations with
independently random divisions into training set and probe set. The
probe set contains 5\% of observed links. Here $L$= 100, 500, 1000,
5000, 10000, 20000.}
{\begin{tabular}{C{1.0cm}|C{1.0cm}C{1.0cm}C{1.0cm}C{1.0cm}C{1.0cm}C{1.0cm}C{1.0cm}C{1.0cm}C{1.0cm}C{1.0cm}}
\hline\hline $L$  &  CN  & Sal & Jac &  S{\o}r &  HPI  &  HDI  &  LHN  &  PA  &  AA  &  RA  \\
\hline
100  &136.7 & 6.603& 9.880& 9.880& 0.271& 14.43& 0.000& 119.3& 147.5& 139.9 \\
500  &48.23 &5.285 &7.946 &7.947& 1.756 &9.405& 0.006 &47.69& 51.89& 48.89\\
1000 &30.97  &6.766 &8.759 &8.759 &0.878& 9.507 &0.035& 30.62 &32.80& 31.29\\
5000 &9.761 &5.427 &5.367& 5.367 &6.574 &4.838 &0.628 &9.483 &9.956&9.973\\
10000&5.397 &4.475& 4.018& 4.018& 4.770 &3.544 &1.383& 5.197 &5.421 &5.453  \\
20000&2.819 &2.749& 2.687 &2.687& 2.791 &2.652& 2.637 &2.786& 2.819 &2.819\\
\hline\hline
\end{tabular}}\label{top-L-popularity}
\end{table*}

\subsection{Results for Cold Probe Set}

We employ the new partition method to prepare the probe set for
experiments. Here we set $D=10$. Thus, we obtain ten different probe
sets $E_{i}^{P}$ ($i=1,\cdots,10$). Clearly, each probe set contains
5\% of observed links, and $\langle \mathrm{pop}\rangle_1<\langle
\mathrm{pop}\rangle_2<\cdots<\langle \mathrm{pop}\rangle_{10}$. The
algorithmic performances on C.elegans network for different probe
sets are shown in Table \ref{CE_auc}. The results for other three
networks are similar.

\begin{table*}[htbp]
\centering
\caption{Algorithmic accuracies on C.elegans for different probe
sets, measured by the AUC value. Sal, Jac and S{\o}r are the abbreviations for
Salton, Jaccard and S{\o}rensen methods respectively. Each value is obtained by averaging
over 100 implementations with independently divisions of training
set and probe set using new partitioning method. The average
popularity of these ten probe sets are shown in the brackets. The
average popularity of the whole set is 523. The highest AUC value in
each row is emphasized in black.}
{\begin{tabular}{L{1.95cm}|C{0.85cm}C{0.9cm}C{0.9cm}C{0.9cm}C{0.9cm}C{0.9cm}C{0.9cm}C{0.9cm}C{0.9cm}C{0.9cm}}
\hline\hline Probe Sets&  CN  & Sal & Jac &  S{\o}r &  HPI  &  HDI  &  LHN  &  PA  &  AA  &  RA  \\
\hline
$E_{1}^{P}$ (57)&0.615&0.724&0.723&0.722&0.713&0.710&\textbf{0.755}&0.247&0.634&0.653 \\
$E_{2}^{P}$ (110)&0.735&0.775&0.780&0.779&0.758&0.778&\textbf{0.787}&0.435&0.756&0.772 \\
$E_{3}^{P}$ (158)&0.754&0.754&0.759&0.759&0.745&0.758&0.748&0.584&0.777&\textbf{0.791}\\
$E_{4}^{P}$ (211)&0.823&0.799&0.806&0.806&0.782&0.805&0.768&0.708&0.842&\textbf{0.849}\\
$E_{5}^{P}$ (283)&0.829&0.777&0.780&0.781&0.773&0.782&0.732&0.806&0.854&\textbf{0.864}\\
$E_{6}^{P}$ (372)&0.867&0.798&0.800&0.800&0.793&0.800&0.726&0.881&\textbf{0.884}&\textbf{0.884}\\
$E_{7}^{P}$ (493)&0.910&0.813&0.807&0.807&0.819&0.797&0.707&\textbf{0.929}&0.921&0.916\\
$E_{8}^{P}$ (650)&0.929&0.818&0.801&0.801&0.846&0.769&0.684&\textbf{0.956}&0.934&0.923\\
$E_{9}^{P}$ (939)&0.943&0.821&0.800&0.800&0.857&0.771&0.669&\textbf{0.980}&0.948&0.939\\
$E_{10}^{P}$ (1987) &0.947&0.814&0.797&0.797&0.848&0.777&0.649&\textbf{0.995}&0.959&0.965\\
\hline\hline
\end{tabular}\label{CE_auc}}
\end{table*}

Compared with other nine indices, LHN has the best performance for
predicting the very unpopular links (the links in $E_{1}^{P}$ and
$E_{2}^{P}$), while has the worst performance on the links in the
probe sets with $P\geq 5$, especially the popular links in $E_8^P$,
$E_9^P$ and $E_{10}^P$. On contrary, PA index gives very good
predictions on the popular links, while extremely bad predictions on
the links with low-degree nodes where the accuracy is even much
lower than the random case. In the middle region where the average
popularity is close to that of the randomly selected probe set, RA
index outperforms others, which is in accordance with the conclusion in
previous studies \cite{ZhouEPJB2009,Lu2010,Liu2010}.

\begin{figure*}[htbp]
\centering
\includegraphics[width=0.4\textwidth]{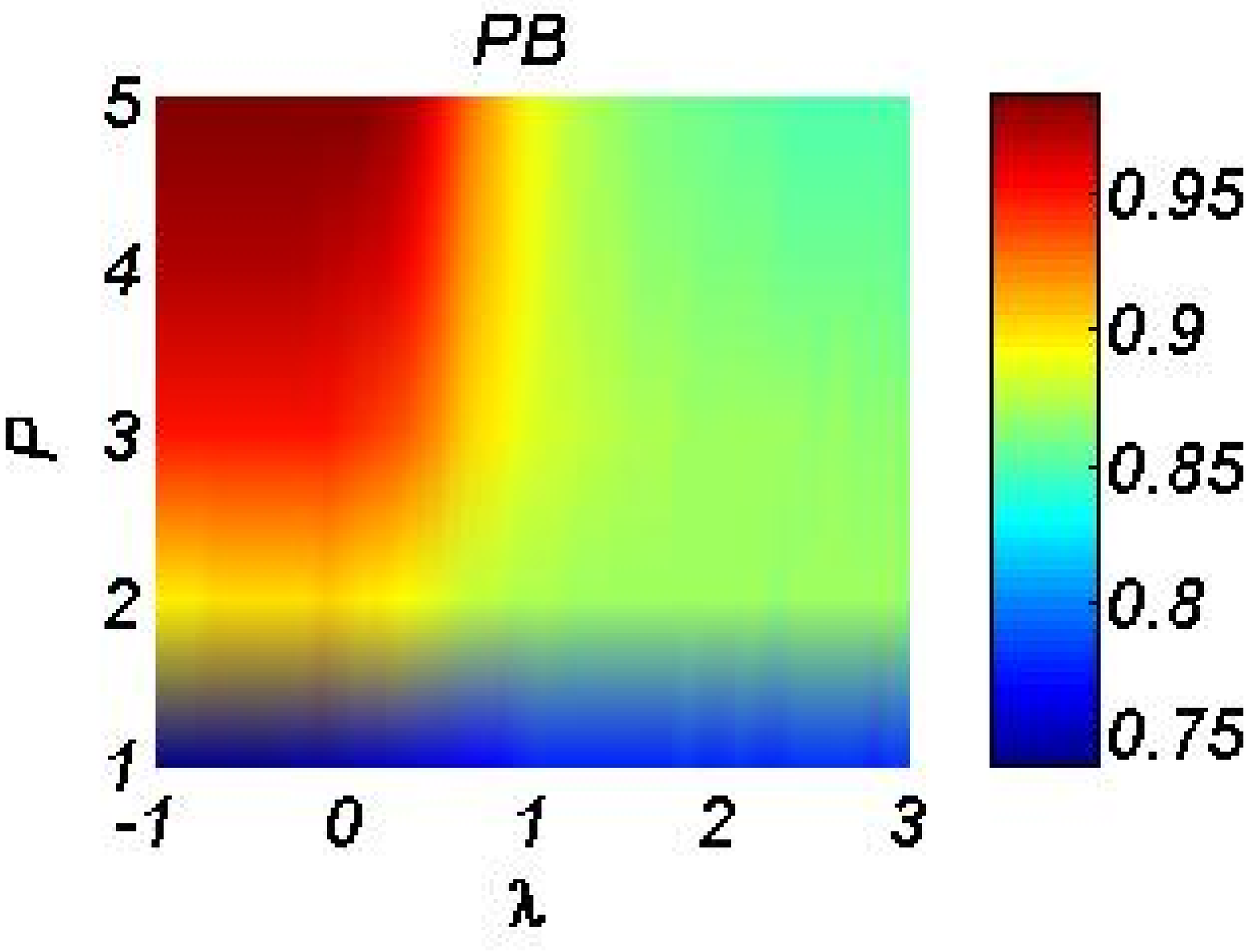}
\includegraphics[width=0.4\textwidth]{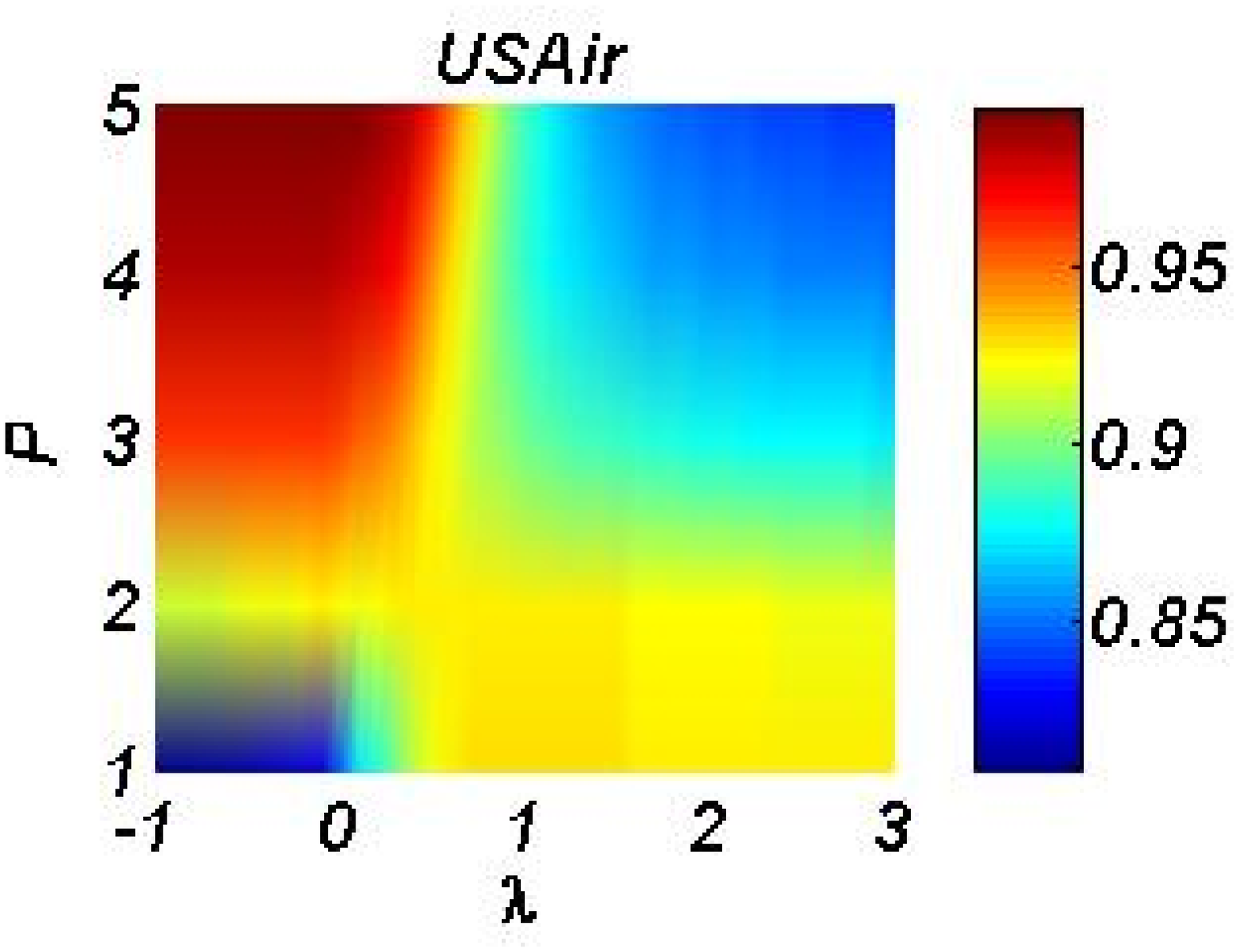}
\includegraphics[width=0.4\textwidth]{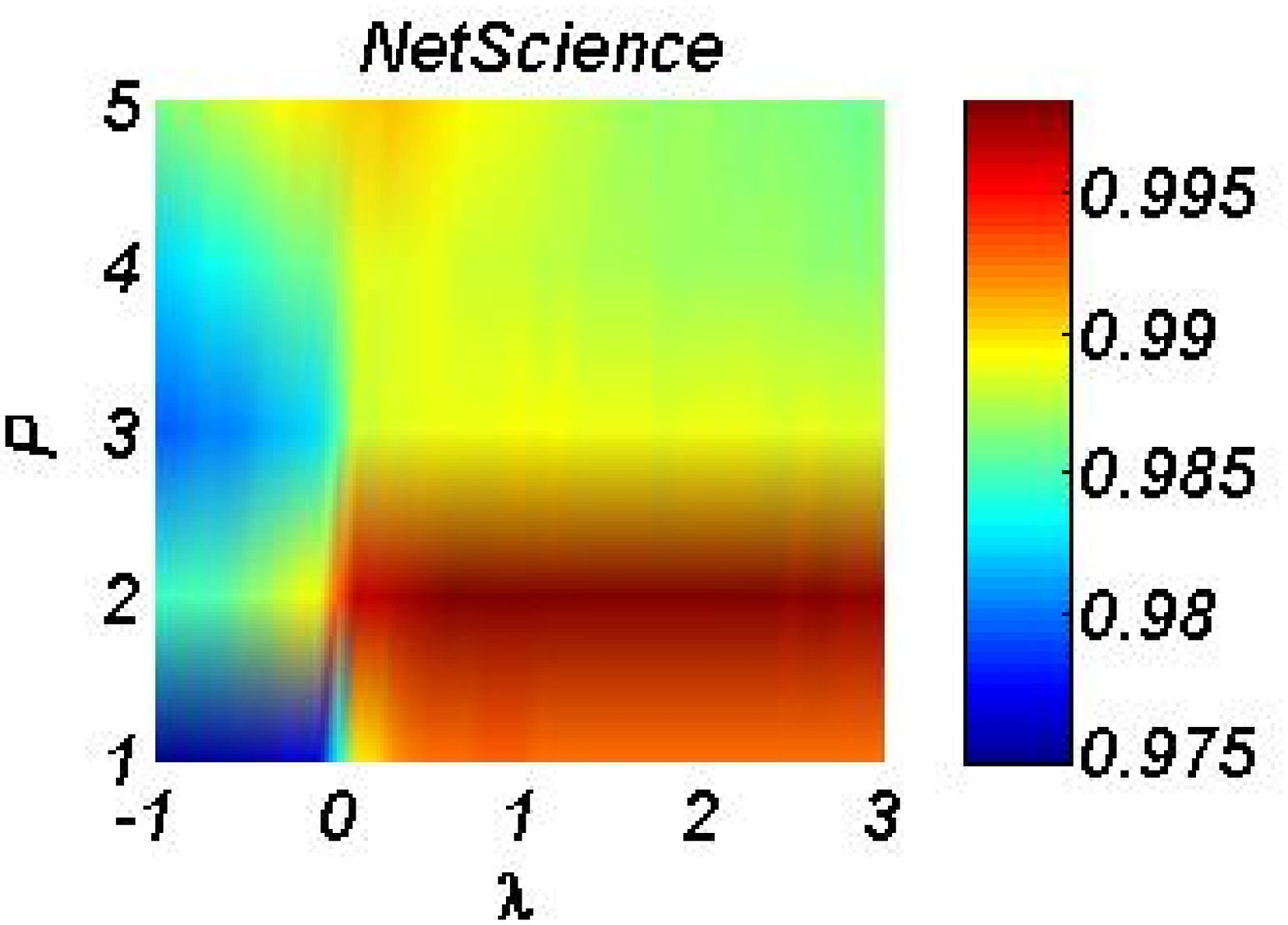}
\includegraphics[width=0.4\textwidth]{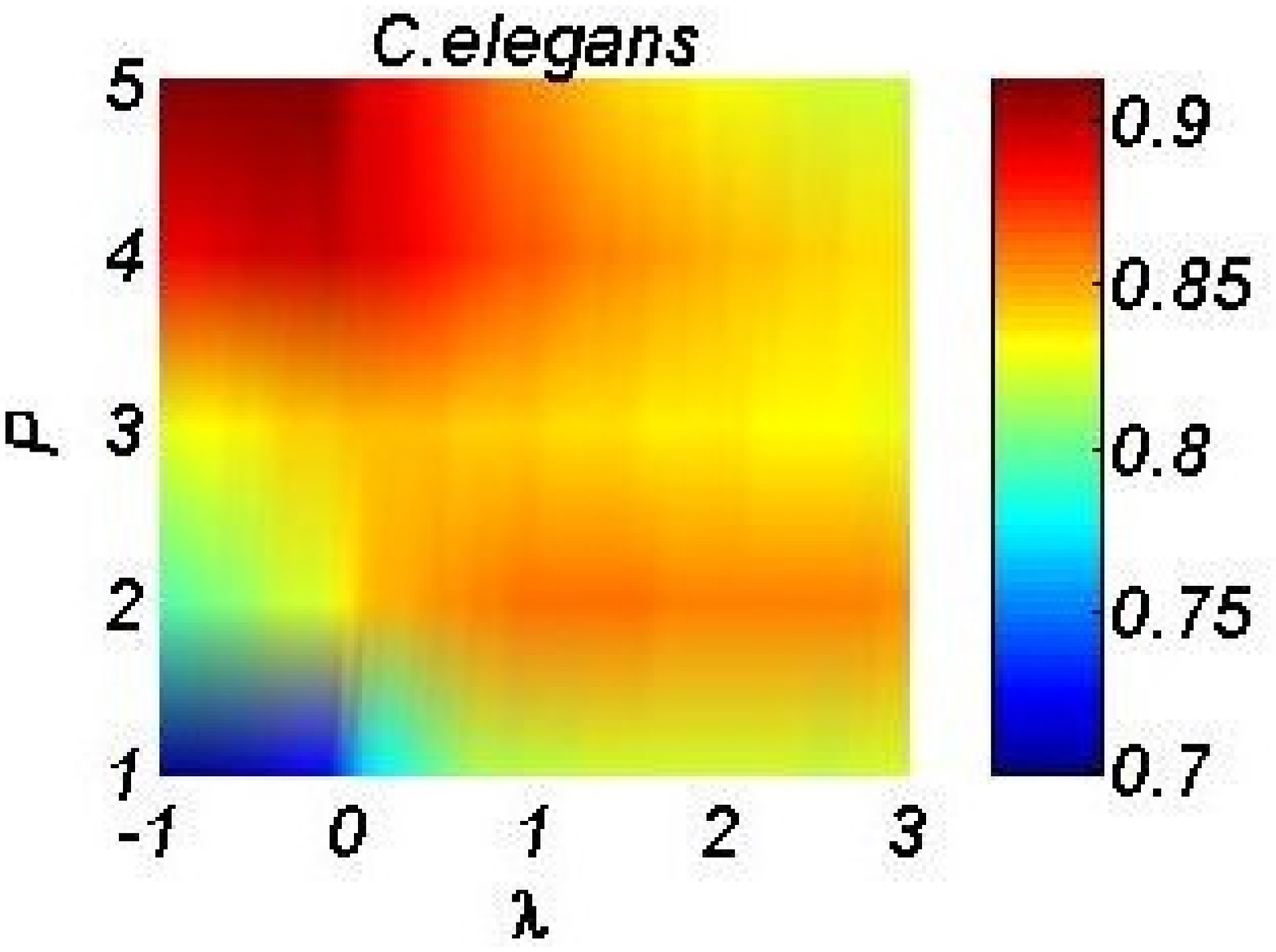}
\caption{The dependence of AUC on different $\lambda$ and $P$. These
results are obtained by averaging over 100 implementations with
independently divisions of training set and probe set (by the new
partitioning method introduced in Sec. II-C and Sec. III-C). Here we
set $D=10$.} \label{heat map}
\end{figure*}

\section{Improved Index}

To design a method for effectively predicting both
popular and unpopular links, we propose a parameter-dependent index,
which is defined as:
\begin{equation}
s_{xy}= \frac{|\Gamma_x\cap \Gamma_y|}{(k_x\times k_y)^{\lambda}},
\end{equation}
where $\lambda$ is a free parameter. This index is also
neighborhood-based and requires only the information of the nearest
neighbors, and thus no extra calculational complexity arises.
Clearly, when $\lambda$ = 0, this index degenerates to CN, and for
the cases $\lambda$ = 0.5 and 1, this index respectively degenerates
to the Salton index and LHN index. Given a network, one can tune
$\lambda$ to find its optimal value subject to the highest accuracy.

\begin{figure*}[htbp]
\centering
\includegraphics[width=3.5in]{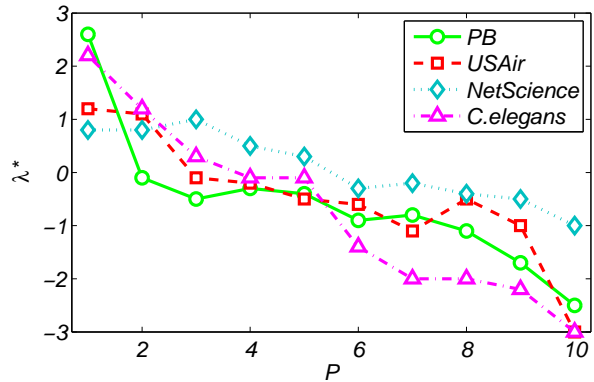}
\caption{The dependence of the optimal parameter $\lambda^*$ on $P$.
The probe set contains 5\% of the known links. Each value is
obtained by averaging the results of 100 independently
implementations.\label{best_coff}}
\end{figure*}

We apply the new index to respectively predict the links in
$E_{i}^{P}$ ($i=1,\cdots,10$). The results of four example networks
in the ($\lambda$,$P$) plane are shown in Fig.~\ref{heat map} where
we focus on the unpopular links ($i=1,\cdots,5$) and $P=i$ means
that $E_{i}^{P}$ is employed as the probe set. The results show that
the optimal $\lambda$ is positive when $P$ is small (i.e., $P$ =
1,2), while it becomes negative for large $P$. For NetScience
$\lambda^*$ becomes negative for $P=6$. The dependence of optimal
parameter $\lambda^*$ on $P$ is shown in Fig.~\ref{best_coff}.
Overall speaking, the optimal parameters $\lambda^*$ of four
networks are negatively correlated with $P$. In other word, the
index with higher (positive) $\lambda$ gives better predictions on
unpopular links, while the index with lower (negative) $ \lambda$ is
good at predicting popular links. For example in C.elegans network,
when $P=1$, namely the probe set is constituted with unpopular
links, the optimal $\lambda^*=2.2$, indicating that to depress the
scores of popular links is a better choice, while for $P=10$, namely
the probe set is constituted with popular links, the optimal
$\lambda^*\approx-3$, which indicates that we had better assign
higher score to the popular links.

The algorithmic accuracies of ten local similarity indices as well
as the proposed index for predicting unpopular links in $E_1^P$
(i.e., $P=1$) are shown in Table \ref{new_auc}. Among the
investigated ten local similarity indices, LHN outperforms others
for predicting unpopular links. However, the proposed index can
improve the accuracy with a proper $\lambda$ which are all positive
for all these four networks. Especially, the improvements on PB and
C.elegans networks are significant, respectively 10.8\% and 8.6\%
compared with LHN.

\begin{table*}[htbp]
\centering
\caption{Comparison of algorithmic accuracy (AUC) on the
cold probe set ($P=1$). NS, CE and PB are the abbreviations for
NetScience, C.elegans and Political Blogs networks respectively. Sal, Jac and S{\o}r are short for
Salton, Jaccard and S{\o}rensen methods. For each network, the probe set contains 5\%
of the known links. Each value is obtained by averaging over 100
independently implementations. The entries corresponding to the
highest accuracies are emphasized in black. For the proposed index,
the AUC values are corresponding to the optimal parameters which are
shown in the brackets.}
{\begin{tabular}{C{1.4cm}|C{0.7cm}C{0.75cm}C{0.75cm}C{0.75cm}C{0.8cm}C{0.75cm}C{0.75cm}C{0.75cm}C{0.75cm}C{0.8cm}C{1.6cm}}
\hline\hline Datasets  &  CN  & Sal & Jac &  S{\o}r &  HPI  &  HDI  &  LHN  &  PA  &  AA  &  RA & New\\
\hline
PB        &0.649 &0.674 &0.664 &0.664 &0.690 &0.662 &0.701 &0.584 &0.656 &0.664 &\textbf{0.777(2.6)}\\
USAir     &0.742 &0.888 &0.881 &0.880 &0.831 &0.875 &0.903 &0.370 &0.792 &0.818 &\textbf{0.928(1.2)}\\
NS        &0.973 &0.991 &0.991 &0.991 &0.988 &0.990 &0.992 &0.095 &0.980 &0.981 &\textbf{0.993(0.8)}\\
CE        &0.615 &0.724 &0.723 &0.723 &0.713 &0.709 &0.756 &0.247 &0.633 &0.654 &\textbf{0.821(2.2)}\\
\hline\hline
\end{tabular}\label{new_auc}}
\end{table*}

\section{Experiments on Real Sampling Methods}

\begin{table*}[htbp]
\caption{Average popularities of missing links corresponding to different sampling methods. 80\% and 90\% means the proportion $|E^T|/|E|$.}
{\begin{tabular}{R{4.22cm}|C{1cm}C{1.3cm}C{1cm}C{1.2cm}C{0.6cm}C{0.9cm}C{0.8cm}C{1cm}}
\hline\hline   &  \multicolumn{2}{c}{PB}  & \multicolumn{2}{c}{USAir} & \multicolumn{2}{c}{NS} &  \multicolumn{2}{c}{CE} \\ \cline{2-9}
& 80\% & 90\% & 80\% & 90\% & 80\% & 90\% & 80\% & 90\%\\
\hline
Acquaintance sampling & 9674.7 & 11543.4&   3115.2 & 3662.8&    84.3&  105.3&   854.7&  1030.9\\
Random-walk sampling & 5361.1  &5220.3  &1658.2 & 1563.6&   49.4 & 41.0 &507.7&  501.1  \\
Path-based sampling  & 4505.1  &4405.3  &1158.4 & 958.1 &18.7 & 18.0    &344.4&  296.6   \\
\hline\hline
\end{tabular}\label{pop_real_samp}}
\end{table*}

\begin{figure*}[htbp]
\centering
\includegraphics[width=0.45\textwidth]{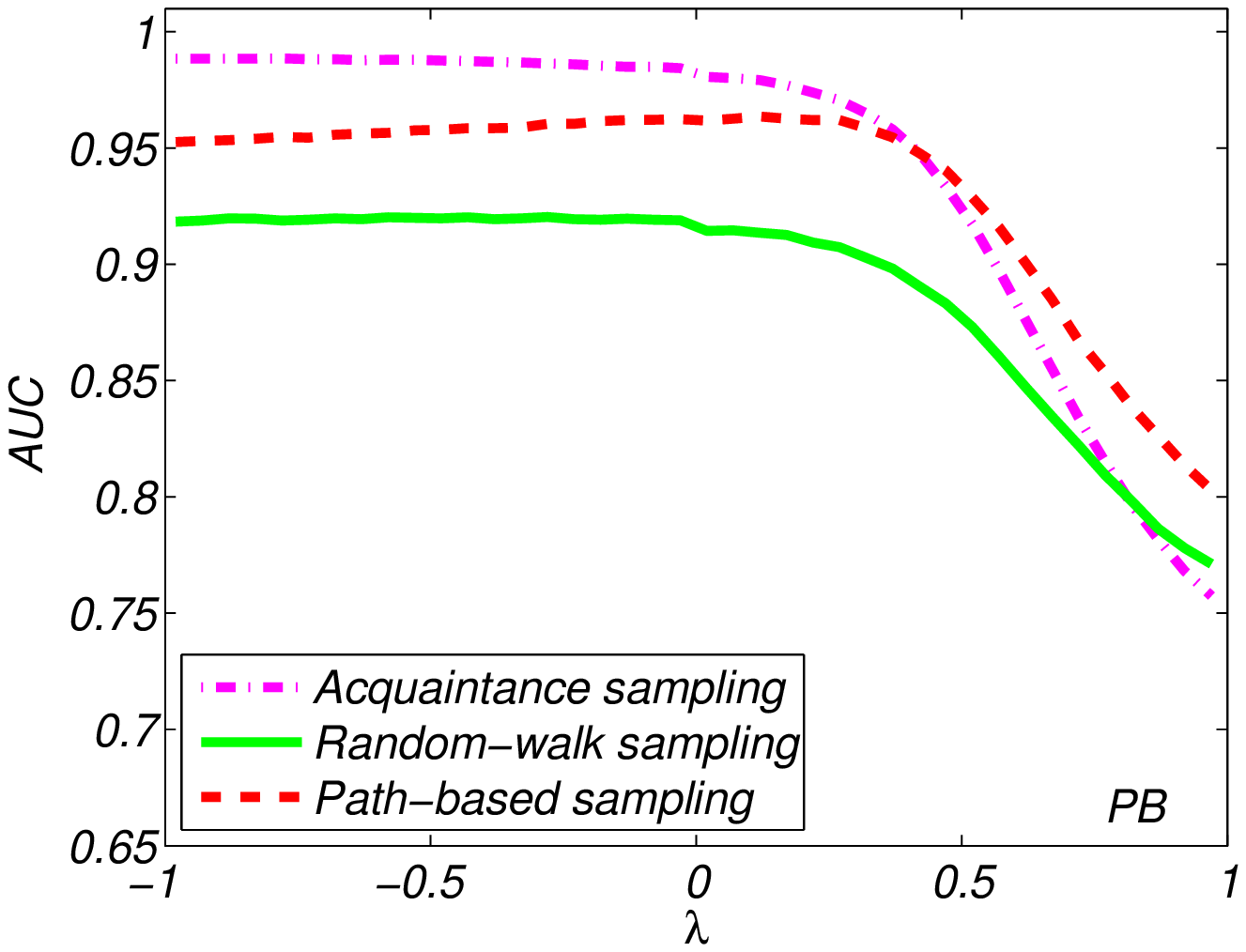}
\includegraphics[width=0.45\textwidth]{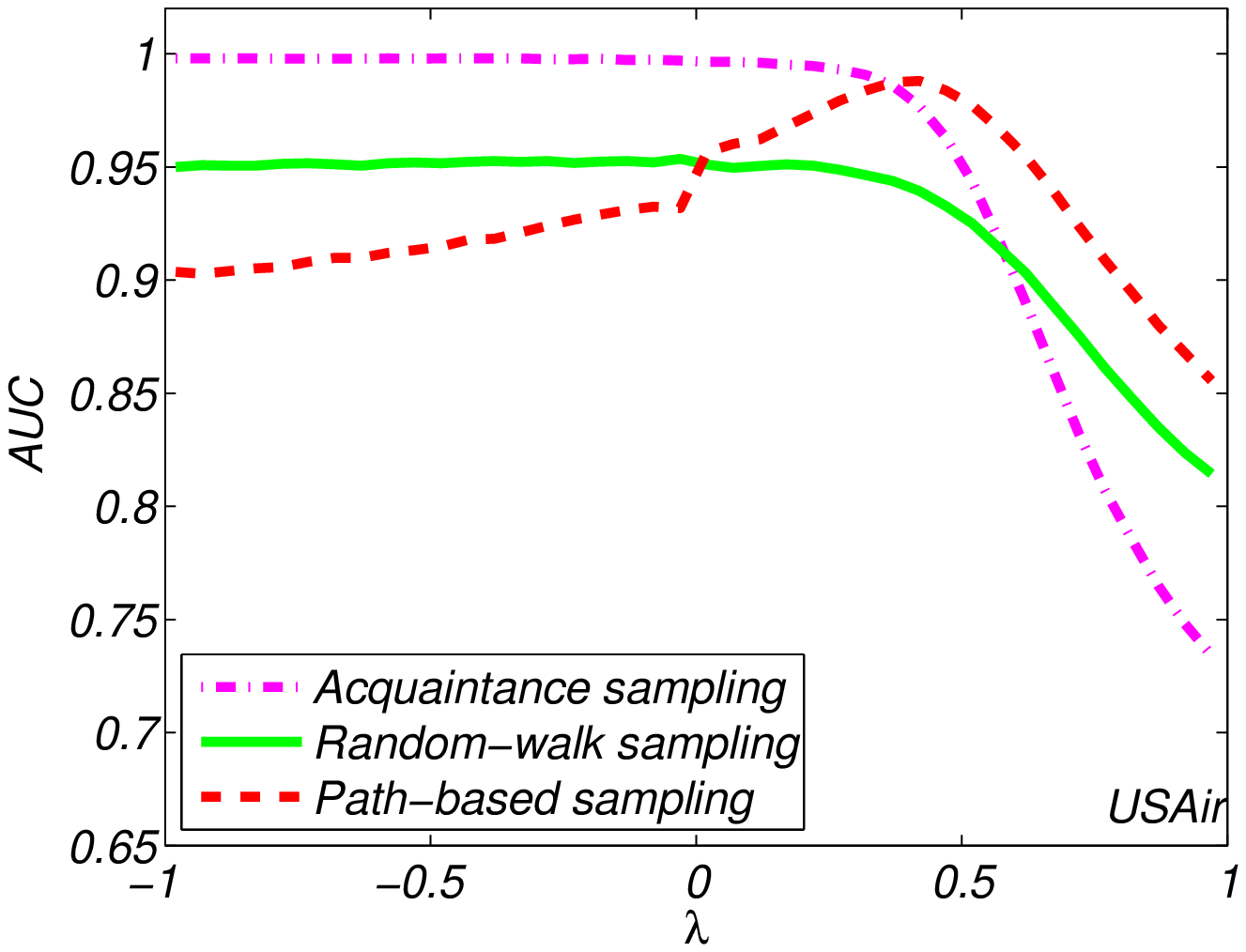}
\includegraphics[width=0.45\textwidth]{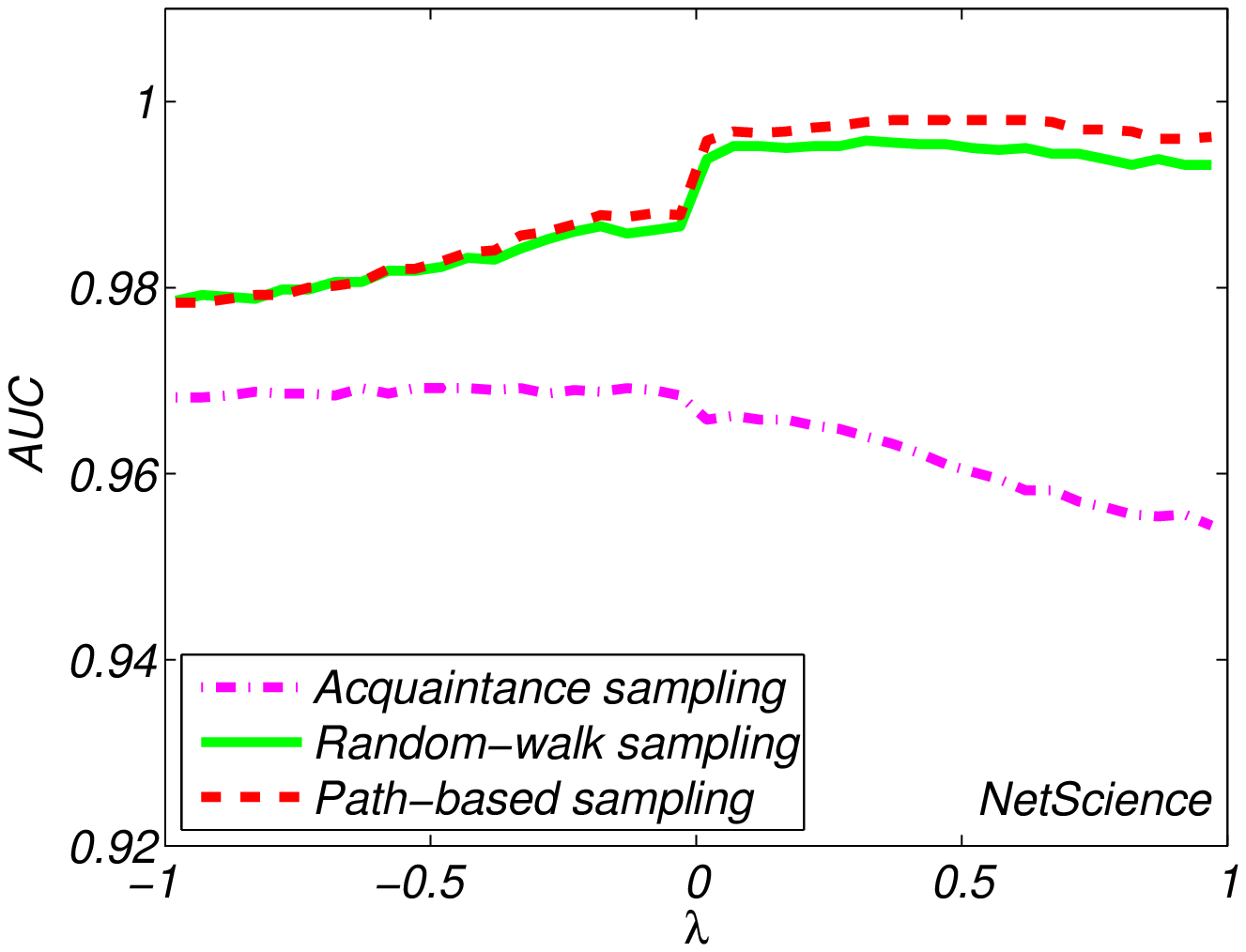}
\includegraphics[width=0.45\textwidth]{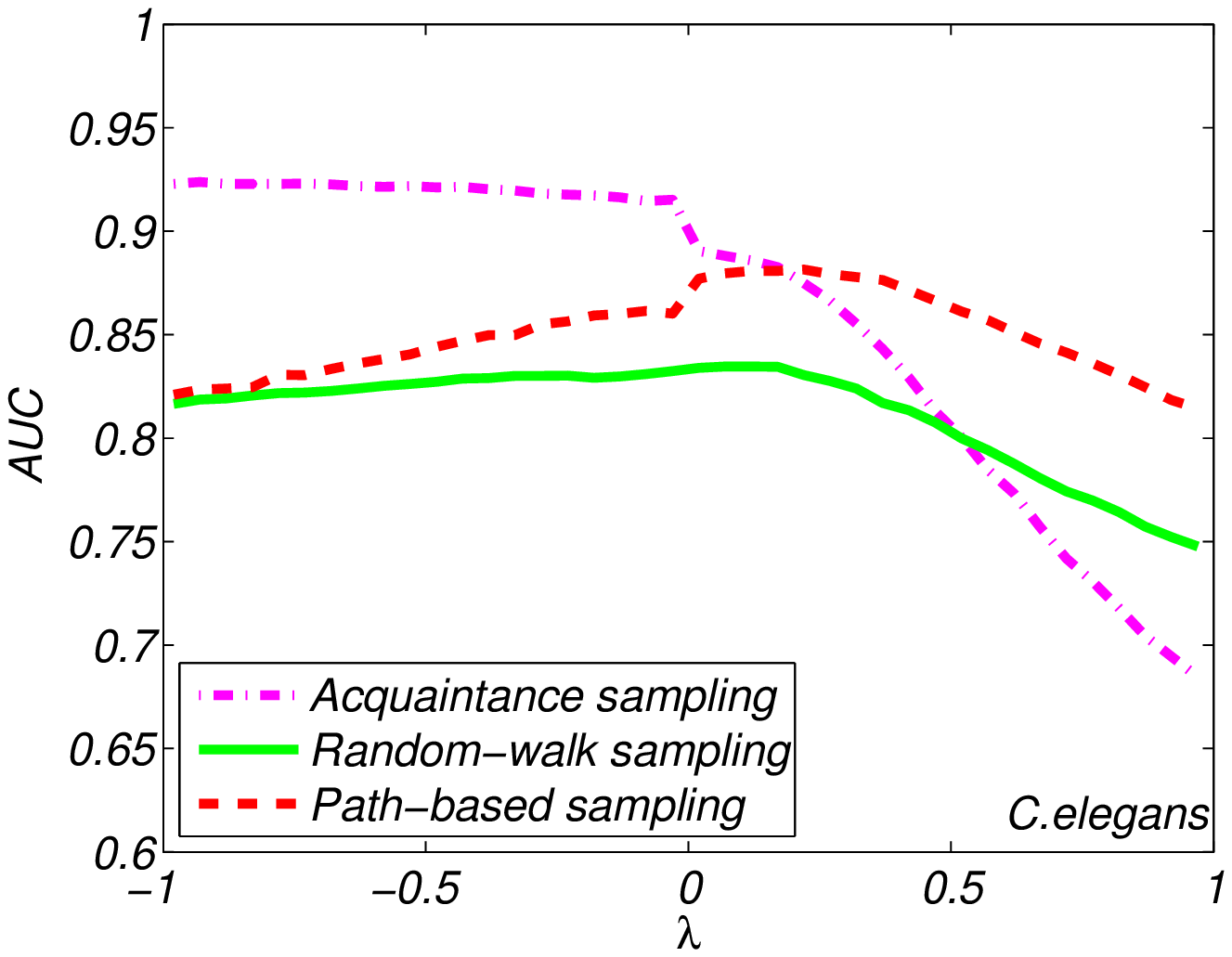}
\caption{Algorithmic performance of the improved index (see Eq. 2) for the three real sampling methods with different values of $\lambda$. The training set contains 90\% of observed links.\label{auc_real_samp}}
\end{figure*}

To connect our study to the real sampled networks, in this section,
we will test the improved index on some real sampling methods.
Firstly, we introduce four mainstream sampling methods as follows.

The first one is called \emph{snowball sampling} (i.e., spider
sampling or breadth first sampling, see Ref. \cite{Biernacki1981}),
which is a non-probability technique and gets widely used in the
studies of World Wide Webs and large-scale social networks. In the
beginning of this method, we randomly select one or a few nodes that
consist of the initial sampled set, and then we crawl all the
neighbors of the nodes in the sampled set, and put them into the
sampled set. This process keeps on until a required number of nodes
are sampled out. Obviously, it is not relevant for the link
prediction problem because this method only leaves missing nodes
rather than missing links.

The second one, called \emph{acquaintance sampling}, is motivated by
epidemic immunization with lack of information \cite{Cohen2003}. In
this method, at each time step, a random link of a randomly selected
node is sampled out (i.e., being put into the training set) until a
required number of links have already been selected. Considering a
link $(x,y)$, if it is not yet sampled out, the probability it will
be selected at this time step is
$\frac{1}{N}(\frac{1}{k_x}+\frac{1}{k_y})$. Although a link with
lower popularity is not necessarily with high
$\frac{1}{k_x}+\frac{1}{k_y}$, statistically speaking, the
probability $\frac{1}{N}(\frac{1}{k_x}+\frac{1}{k_y})$ is negatively
correlated with the popularity $(k_x-1)(k_y-1)$. To our knowledge,
this method is a very special method where unpopular links are more
likely to be sampled out yet the popular links consist of the probe
set.

The third one is named \emph{random-walk sampling}
\cite{Lovasz1993}. A simple way adopted is as follows: (i)
initialize a particle on a randomly selected node; (ii) this
particle jumps to a randomly selected neighbor and the corresponding
link will be added into the training set (i.e., sampled out); (iii)
repeat (ii) until a certain number of links have been sampled out,
and the rest links compose the probe set. It is well-known that the
distribution of visiting frequency of a random walker on a connected
network will soon converges to the degree distribution, namely the
probability at a certain time step the random walker locate in a
node $x$, say $\psi(x)$, is equal to $\frac{k_x}{2M}$, where
$2M=\sum_yk_y$ serves as a normalization factor. Considering a link
$(x,y)$, if it is not yet sampled out, the probability it will be
selected at this time step is
$\psi(x)\frac{1}{k_x}+\psi(y)\frac{1}{k_y}=\frac{1}{M}$. That is to
say, the average popularity of links in the probe set is
approximately the same to that from the random sampling (we have
checked it by simulation). However, the random-walk sampling is not
the same as random sampling, for example, the sampled network from
the former is always connected yet the one from the latter may
contain several components.

The last one is called path-based sampling, which has been applied
in extracting the topology of Internet at router level
(http://www.routerviews.org). Indeed, this method tracks the
transmission of information packets in the Internet, and a link
passed by more packets has higher probability to be sampled out. To
simulate this process, at each time step, we randomly select a
starting point and an end, and we assume that a packet will go from
the starting point to the end through a randomly selected shortest
path. After a sufficiently large number of time steps, a link with
more than a threshold, $N_T$, packets will be put into the training
set while others compose the probe set. Here for simplicity, we set
$N_T=1$. Under this method, the links with high betweenness
centrality (betweenness centrality quantifies the traffic load of a
link, depending on the routing strategy of packet transmission
\cite{Yan2006}) are favored. Since the popularity of a link is
strongly positively correlated with its betweenness centrality on
shortest-path routing, the average popularity of links in the probe
set is lower than that of the random sampling.

\begin{figure*}[htbp]
\centering
\includegraphics[width=3.5in]{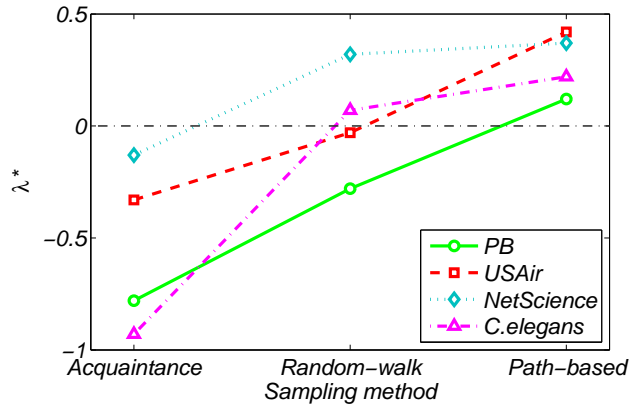}
\caption{The optimal values of $\lambda$ for the three sampling
methods. The dash line denotes $\lambda =0$ as an eye guidance.
\label{optimal_lambda}}
\end{figure*}

The average popularities of missing links corresponding to different
sampling methods are shown in Table 6. Agreeing with our analysis,
the average popularities of the links in the probe set obeys the
inequality $\langle \texttt{pop}\rangle_{acquaintance} > \langle
\texttt{pop}\rangle_{random-walk} > \langle
\texttt{pop}\rangle_{path-based}$. Figure 4 reports the algorithmic
performance (measured by the AUC value) for different sampling
methods and different $\lambda$. Very clearly, aiming to predict
popular links (e.g., under acquaintance sampling) $\lambda^*$ is
negative while to predict cold links, $\lambda^*$ is positive. In
fact, there is strongly negative correlation between the average
popularity of links in the probe set and the optimal value of
lambda. As shown in Figure 5, where we order the three sampling
methods with decreasing average popularity of missing links and thus
a positive correlation is observed.

\section{Conclusion and discussion}

To our knowledge, in the previous studies on link prediction
\cite{LPsurvey}, the data sets are always divided in a random
manner. Inspired by the in-depth thought about the features of
missing links, this article challenges such a straightforward
method. Applying a simple measure on link popularity, we propose a
method to sample less popular links for probe set. Experimental
analysis shows a surprising result that the LHN index performs the
best, although it was known to be one of the worst indices if the
probe set is a random sampling of all links. We propose a similarity
index with a free parameter $\lambda$, by tuning which this index
can degenerate to the Common Neighbor index, the Salton index and
the LHN index. The optimal value of $\lambda$ monotonously depends
on the average link popularity of probe set. We further test this
index on three real sampling methods. Agreeing with the main results
from theoretical analysis, the optimal value of $\lambda$ increases
with the decreasing of the average popularity of links in the probe
set. Again, the improved index in a well-tuned range can outperform
others under real sampling methods.

Notice that, the main contribution of this article does not lie on
the proposed index. Instead, the significance of this work is to
raise the serious question about how to properly determine the probe
set. To us, this is a very important yet completely ignored problem
in information filtering. The reconsideration of dataset division
will largely change the understanding and thus the design of
algorithms in information filtering (also relevant to the so-called
recommendation problem \cite{Zhou2010}). As a start point, we give a
naive method and a preliminary analysis, which is of course far from
an satisfied answer to the question. In fact, we think in-depth
understanding of real sampling methods may shed light on this issue.

\section*{ACKNOWLEDEMENTS}
This work is partially supported by the National Natural Science
Foundation of China under Grant Nos. 11075031 and 10635040, the
Swiss National Science Foundation under Grant No. 200020-132253, and
the Fundamental Research Funds for the Central Universities.

\end{document}